\documentclass[journal]{IEEEtran}
\usepackage{bm}
\usepackage{graphicx}
\usepackage{multirow}
\usepackage{amsmath,amssymb,amsfonts}
\usepackage{array}
\usepackage{color}
\usepackage{doi}

\ifCLASSINFOpdf
\else
\fi
\begin{document}

\title{Cascade LSTM Based Visual-Inertial Navigation for Magnetic Levitation Haptic Interaction}

\author{Qianqian Tong,
        Xiaosa Li,
        Kai Lin,
        Caizi Li,
        Weixin Si,
        Zhiyong Yuan
% <-this % stops a space
%\thanks{Jeungeun Song is with The University of British Columbia, Email: jeungeunsong@ieee.org}% <-this % stops a space
\thanks{Q. Tong, X. Li, C. Li, and Z. Yuan are with Wuhan University, China. Email: zhiyongyuan@whu.edu.cn}% <-this % stops a space
\thanks{K. Lin is with Dalian University of Technology, China}
\thanks{W. Si is with Guangdong Provincial Key Laboratory of Machine Vision and Virtual Reality Technology, Shenzhen Institutes of Advanced Technology, Chinese Academy of Sciences, Shenzhen, China. Email: wxsics@gmail.com}
%\thanks{Weixin Si and Zhiyong Yuan are both the corresponding authors.}
}

%\markboth{Under Review,~Vol.~xx, No.~xx, Month~20xx}%

\maketitle

\begin{abstract}
Haptic feedback is crucial to immersive experience in virtual and augmented reality applications. The existing promising magnetic levitation (maglev) haptic devices have advantages of none mechanical friction and low inertia. However, their performance is limited by the navigation approach, which mainly results from the challenge that it is difficult to obtain high precision, high frequency and good stability with lightweight design at the same time. In this study, we reformulate visual-inertial navigation as a regression problem, and adopt deep learning to perform fusion navigation for maglev haptic interaction. A cascade LSTM based $\theta $-increment learning method is first proposed to progressively learn the increments of target variables. Two cascade LSTM networks are then constructed to respectively estimate the increments of position and orientation which are pipelined to accomplish visual-inertial fusion navigation. Additionally, we set up a maglev haptic platform as the system testbed.
Experimental results show that our cascade LSTM based visual-inertial fusion navigation approach can reach 200Hz while maintaining high-precision (the mean absolute error of the position and orientation is less than 1mm and 0.02$^{\circ }$, respectively) navigation for a maglev haptic interactive deformation application.
\end{abstract}


% \fi
\IEEEpeerreviewmaketitle

\section{Introduction}

The recent development of virtual reality (VR) and augmented reality (AR) has facilitated the advancement of related applications, such as surgical procedures, teaching-learning system, marketing research, and interactive recreation \cite{chen2018wearable,elbamby2018toward}. In these applications, haptic sensation is an essential component of users' immersive interaction experience.
%when interacting in virtual or augmented reality.
%To provide haptic interaction with low static friction and low inertia,
Berkelman \emph{et al.} \cite{berkelman2012co,pedram2017torque} developed a maglev haptic interface which provided haptic feedback via a penhandle or fingertip probe. Besides, a novel maglev haptic device with an adjustable coil configuration was deployed, and it can provide haptic feedback in a natural manner \cite{tong2016novel,tong2017magnetic}. For these maglev haptic devices, the position and orientation of their magnetic stylus\verb|/|probe are firstly obtained to navigate users' interaction actions. High-precision and high-speed navigation helps to capture subtle changes in users' actions.
%, which is the prerequisite for providing immersive haptic experience. However, if users' interaction operation cannot be acquired accurately and quickly,
Conversely, if users' actions cannot be acquired accurately and quickly, the haptic experience will be distorted. Therefore, the navigation performance is crucial to providing immersive haptic feedback.

In the study of Berkelman \emph{et al.} \cite{berkelman2012co,pedram2017torque}, an Optotrak Certus 6 degrees-of-freedom (DOF) optical motion tracker (Northern
Digital Inc.) provided real-time position and orientation feedback for their maglev haptic platform. Infrared LEDs with no wired connection were mounted on the back end of their user's probe. However, the infrared LEDs will be sheltered by each other when the probe is tilted at a large angle, leading to the loss of location information. Besides, the design of their tracking module is somewhat cumbersome because of the additional mass and bulk of the battery and electronics required in wireless mode.
%Although their position sensing system possesses high resolution, its effective workspace is limited.

Tong \emph{et al.} \cite{tong2016novel,tong2017magnetic} designed a magnetic stylus consisting of several small rods and red markers were embedded in connections between these small rods. A visual module with two RGB cameras was utilized to track red markers in the magnetic stylus for obtaining user's interaction actions. Although this visual module has advantages of high precision, light weight and low cost, the positioning frequency is limited by cameras' low acquisition frequency while maintaining high precision, which will affect the resolution of haptic perception. Besides, there also exists occlusion problem in this visual module.

From the above observations, three challenges should be addressed for existing navigation methods to accomplish high quality navigation for maglev haptic interaction. Firstly, how to maintain high positioning frequency while providing high precision. Secondly, how to improve the stability and robustness when the occlusion problem occurs, i.e., how to tackle the possible occlusion problem when the probe is tilted at a large angle or when users operate several probes in the operation workspace at the same time. Thirdly, how to design a lightweight and cost-effective navigation module while addressing the above two challenges.

In this work, we resort to the fusion navigation scheme which is capable of taking advantages of different navigation methods to overcome the aforementioned challenges. Considering that inertial navigation has advantages of high sampling frequency and good stability, and these characteristics are complementary to the high precision of visual navigation, we adopt inertial measurement units (IMUs) to aid the visual module thus to advance the navigation performance for maglev haptic interaction.

\begin{figure*}[htbp]
  \centering
  \includegraphics[width=13.5cm]{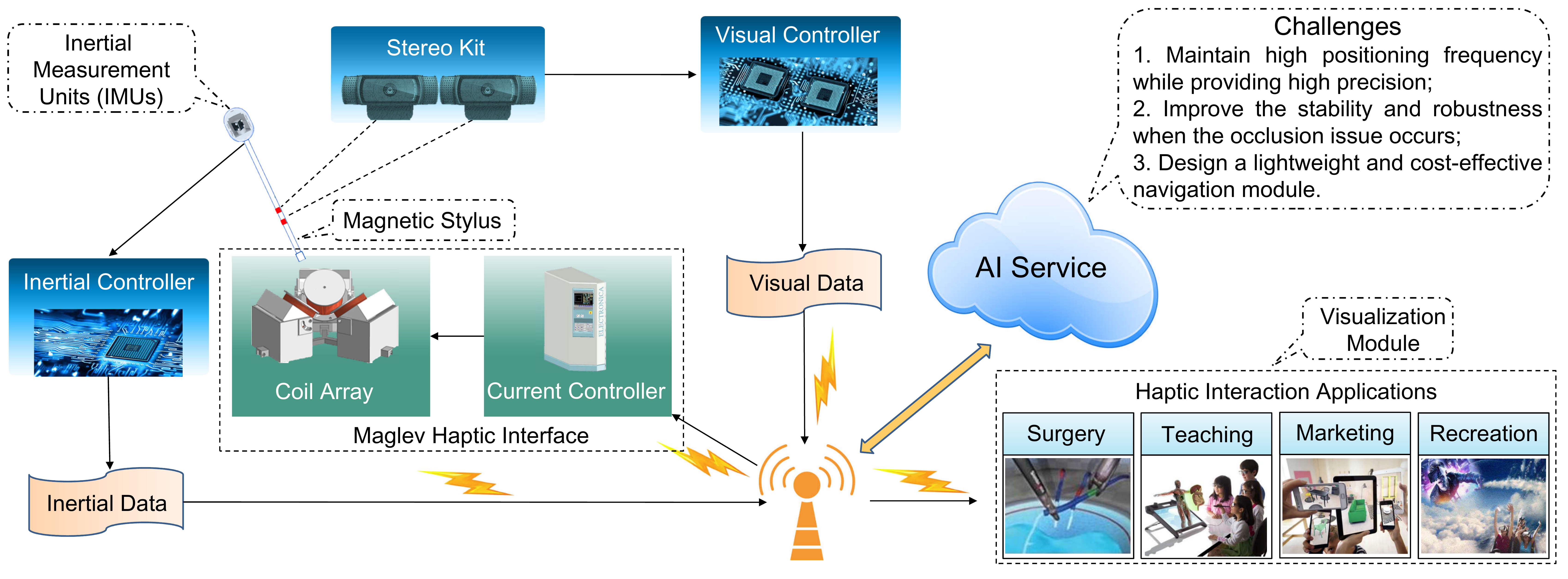}\\
  \caption{An overall design for our maglev haptic interaction system.}\label{fig1}
\end{figure*}

Several researchers have explored many kinds of visual-inertial (VI) navigation methods \cite{weiss2011real,mourikis2007multi,leutenegger2015keyframe,qin2018vins}. These methods can be tightly-coupled or loosely-coupled according to the condition whether image features are part of the state vector. Although tightly-coupled fusion methods can provide long-term, high-precision navigation, they usually involve filter update based on a certain constraint or an optimization problem, leading to low positioning frequency. Loosely-coupled methods maintain the integrity of visual module and IMUs, which is convenient for their independent optimization.
%In order to improve the positioning speed, we adopt the loosely-coupled fusion mode. Our work aims at accomplishing a kind of accurate, high-frequency, robust and real-time navigation by fusing stereoscopic tracking module with inertial navigation.

Inspired by the recent success of deep learning techniques \cite{chen2018label,chen2017deep}, especially the great advancement of the long short-term memory (LSTM) architecture for recurrent neural networks (RNNs) \cite{greff2017lstm}, we regard visual-inertial navigation as a regression problem and employ a deep learning approach to perform visual-inertial navigation. In our work, visual and inertial modules are calibrated with reference to the fusion navigation coordinate system, respectively. To accomplish high-speed and robust navigation, we present a cascade LSTM based $\theta $-increment learning method and construct two cascade LSTM networks to estimate increments of position and orientation, respectively. The estimated increments are then pipelined to calculate the position and orientation of the moving object. Finally, this location information is used for navigating maglev haptic interaction applications.

The main contributions of this paper are as follows:

$\bullet$ We reformulate visual-inertial navigation as a regression problem and propose a novel visual-inertial fusion navigation approach based on deep learning for maglev haptic interaction. This approach is excellent in real-time performance while maintaining high precision.

$\bullet$ We present cascade LSTM based $\theta $-increment learning for accomplishing visual-inertial navigation, and two cascade LSTM networks are constructed to estimate increments of position and orientation. The accuracy of our cascade LSTM based navigation approach is verified by experimental results.

$\bullet$ Our cascade LSTM based visual-inertial navigation approach is lightweight, cost-effective, as well as robust to the occlusion problem. Furthermore, it can be extended to other applications other than being utilized in the maglev haptic interaction application mentioned in this work.

%$\bullet$ Our visual-inertial fusion navigation method is robust to the occlusion issue by directly learning the relationship between the position and orientation of the stylus and the inertial measurement unit.
%
%$\bullet$ The presented visual-inertial fusion navigation module has advantages of lightweight and cost-effective, and it can be easily extended to other applications other than navigating the magnetic levitation touch interaction mentioned in this article.

The remainder of this article is organized as follows: In Section II, the problem statement and system architecture are introduced. The proposed cascade LSTM based visual-inertial navigation scheme is given in Section III. In Section IV, the system testbed and experimental results are presented. Finally, conclusions are given in Section V.

\section{Problem statement and system overview}

In this section, we first introduce the problem statement about the navigation method in maglev haptic interaction applications. Then, we provide the system overview, as shown in Fig.~\ref{fig1}.

\subsection{Problem statement}

Visual navigation is usually used to capture users' actions in maglev haptic interaction applications \cite{berkelman2012co,pedram2017torque,tong2017magnetic} because of its high-precision. However, its output frequency of the position and orientation is low and unstable due to its low sampling frequency and environmental conditions of cameras. Besides, the navigation performance will be affected if markers are out of cameras' field of view.
%Inertial navigation measures the acceleration and angular rate of {\color{red}the moving object} using IMU which includes a triaxial accelerometer and a gyroscope, and then works out its pose, speed, localization and so on by means of transformation and integration.

Inertial navigation system (INS) is all-weather, and it can work in various environments and export in-motion data at a high frequency.
To obtain the position and orientation of the moving object by INS, inertial navigation coordinates should be selected firstly. IMU including an accelerometer and a gyroscope is connected to the moving object to gather a raw acceleration $\bm{a}$ and an angular rate $\bm{\omega }$ in the inertial navigation coordinates. The orientation can be determined by transforming the quaternion updated by Runge-Kutta Act method particularly to orientation angles (pitch, roll, yaw). The raw acceleration $\bm{a}$ should be converted to the motion acceleration $\bm{a}_n$ in the inertial navigation coordinates before the integration for the position. The position can be simply obtained by the integration of acceleration. Note that the integration operation is usually done in the frequency domain through the fast Fourier transformation to reduce the error caused by biases and high-frequency noises. Though some measures are token to decrease the influence of biases and noises, computational errors still can not be eliminated completely. What's worse, the error will accumulate over time.

Considering that characteristics of visual and inertial navigation are complementary, cameras and IMUs are usually fused to acquire state estimations. Weiss \emph{et al.} \cite{weiss2011real} coupled the visual framework and IMU loosely. They treated the visual framework as a black box, and showed how to detect failures and estimated drifts in it. Mourikis \emph{et al.} \cite{mourikis2007multi} put forward a multi-state constraint Kalman Filter (MSCKF) algorithm, which performed an Extended Kalman
Filter (EKF) update based on geometric constraints. Apart from filter-based methods, there are also optimization-based methods such as keyframe-based visual-inertial SLAM (OKVIS) using nonlinear optimization proposed by Leutenegger \emph{et al.} \cite{leutenegger2015keyframe}. Besides, VINS-Mono presented by Qin \emph{et al.} \cite{qin2018vins} is a nonlinear-optimization-based sliding window estimator using pre-integrated IMU factors.
%In the study of Li \emph{et al.} \cite{li2013high}, estimator initialization and failure recovery should be finished firstly. Then, they used a nonlinear optimization-based method to obtain high accuracy visual-inertial navigation after finishing the pre-integration of IMU measurements. Their estimator was robust. Through pre-integration and estimator initialization, we can improve our method in advance of real-time navigation.

%It should be noted that the aforementioned traditional visual-inertial navigation methods involves filter update based on a certain constraint or an optimization problem, which needs complex calculations, making it hard to reach high fame rates.

The aforementioned visual-inertial navigation methods have been applied to state estimation problems in a variety of fields, such as autonomous vehicles and flying robots \cite{Delmerico2018A}. However, these methods still have many drawbacks. Specifically, the inertial and visual processing frequency of the method in \cite{weiss2011real} are only 75Hz and 25Hz, respectively. Although MSCKF \cite{mourikis2007multi} is robust and memory-efficient, its per-frame processing time is also long and its accuracy is low. Moreover, the accuracy of OKVIS \cite{leutenegger2015keyframe} and VINS-Mono \cite{qin2018vins} is relatively high, but this achievement greatly sacrifices computational resources, leading low processing frequency. What's worse, the above methods can only achieve the accuracy of decimeter. Therefore, these existing visual-inertial navigation methods are not suitable for maglev haptic interaction which needs high precision and high frequency for immersive interaction experience. In our work, we takes advantages of visual and inertial navigation by reformulating visual-inertial navigation as a regression problem using deep leaning, aiming at improving the navigation frequency while maintaining high precision.

Although VINet \cite{clark2017vinet} similarly regarded the visual-inertial odometry as a sequence-to-sequence regression problem, its fusion navigation frequency is limited by the low-frequency data stream, such as the visual or ground truth data stream.
%Due to the different frame rates of visual and inertial sensors, the data streams used for fusion navigation are multi-rate. To tackle this issue,
In this work, we present a cascade LSTM based $\theta $-increment learning method to progressively learn increments of position and orientation at a small time step. Suppose that the time step of the ground truth is $T$, and the time step of our $\theta $-increment learning method for navigation estimation is $t$. Note that $t$ can be small than $T$ in our study. From this perspective, our $\theta $-increment learning based visual-inertial fusion navigation method can reach higher frequency than that of the ground truth. The implementation details of the presented cascade LSTM $\theta $-increment learning method and our visual-inertial fusion navigation approach will be introduced in Section \ref{caslstm}.

\subsection{System overview}

As shown in Fig.~\ref{fig1}, the maglev haptic interaction system is composed of a visual acquisition unit (stereo kit), two IMUs, a visual controller, an inertial controller, a haptic feedback interface, a current controller, an AI service and a visualization module, etc. In this study, the East-North-Up coordinate system is chosen as the fusion navigation coordinate system, and the navigation task is to capture the position and orientation of a moving object relative to the selected coordinate system. IMUs are fixed on the back end of the magnetic stylus. The visual and inertial controller are connected to a router by the Ethernet connector and send data collected by sensors to the AI service under the same LAN.
%One is used to gather the raw acceleration and angular rate. The other one is used for providing orientation angles.

When an operator uses the magnetic stylus to interact with virtual scenes, the stereo kit with two cameras acquires RGB images and IMUs obtain acceleration and angular rate. Visual controller is used to calculate the position of the magnetic stylus and one inertial controller is used to calculate its orientation. The calculated position and orientation, and the collected acceleration and angular rate are used to estimate the final position and orientation of the magnetic stylus with high frequency through the AI service.

After the AI service calculates out the position and orientation of the magnetic stylus by using the proposed cascade LSTM based visual-inertial navigation approach, it sends the navigation information to the visualization module. The visualization module performs collision detection between the virtual stylus and virtual objects, and meanwhile computes the feedback force to be exerted on the magnetic stylus. Then, the current to be loaded for each coil in the coil array of the maglev haptic interface is calculated according to the calculated feedback force. The current controller intelligently adjusts the current of each coil \cite{tong2017magnetic}, making the coil array generate effective magnetic field corresponding to the interactive process. Finally, the magnetic stylus receives the same force as the virtual stylus and transmits it to the operator.

%\IEEEpubidadjcol

\section{Cascade LSTM based VI Navigation}  \label{caslstm}

In this section, we first give an overview of the presented cascade LSTM based $\theta $-increment learning method. Then, we construct two cascade LSTM networks using the $\theta $-increment learning method to estimate increments of position and orientation. Finally, our visual-inertial navigation approach based on cascade LSTM is described.

\subsection{Cascade LSTM based $\theta $-Increment Learning} \label{casunit}

\begin{figure*}[htbp]
  \centering
  \includegraphics[width=12.7cm]{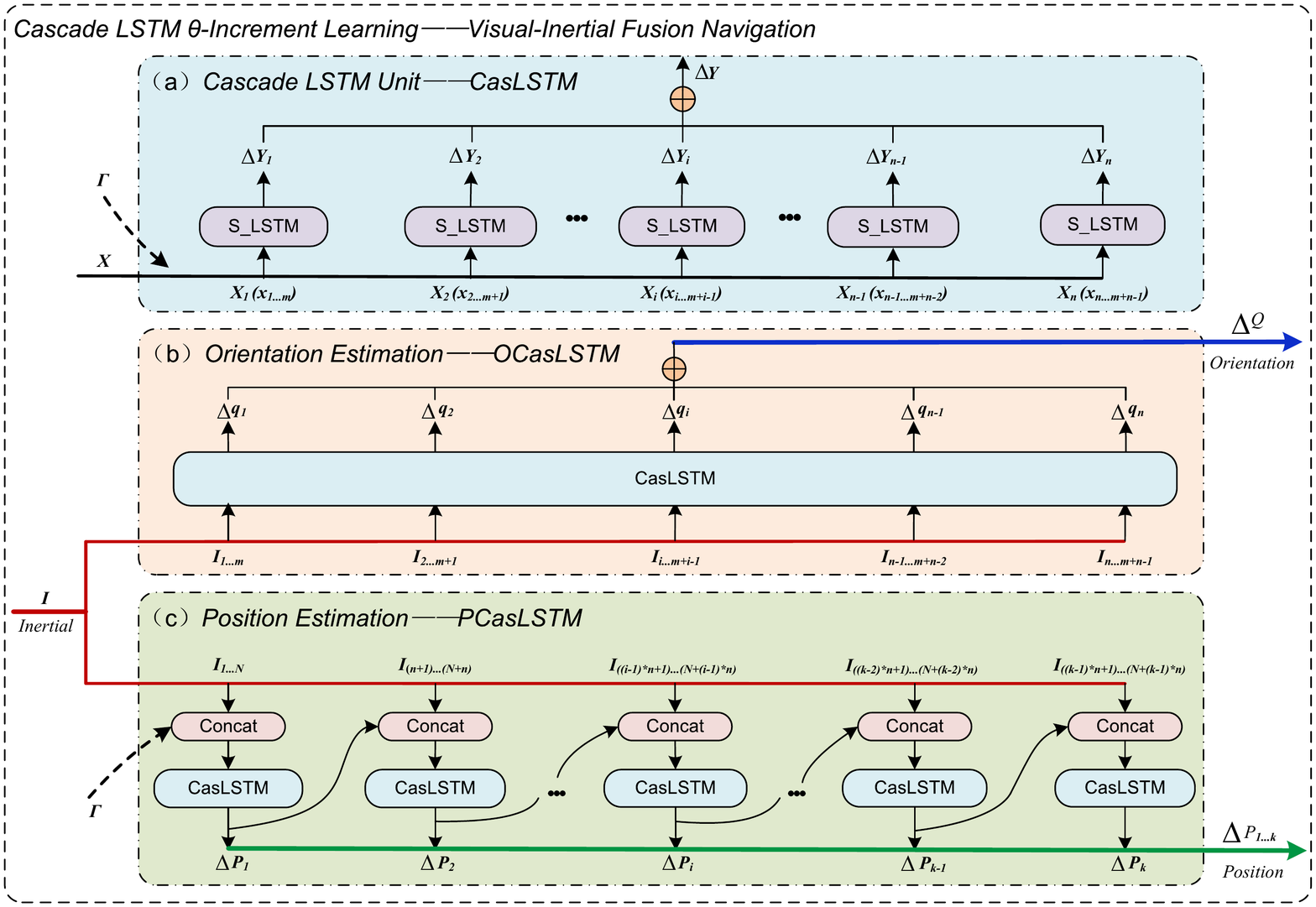}\\
  \caption{The architecture of cascade LSTM network for visual-inertial navigation.}\label{fig2}
\end{figure*}

Due to different sampling frequencies of visual and inertial sensors, data streams used for navigation are multi-rate, and the frequency of visual data is lower than that of inertial data. It is challenging to realize visual-inertial navigation with high frequency using these multi-rate data for deep learning models. To tackle this issue, we present a $\theta $-increment learning method by constructing a cascade LSTM network unit to progressively learn increments of target variables, as shown in Fig.~\ref{fig2}(a).

Giving an input $\bm{X}=(\bm{x}_1,\bm{x}_2,\cdots,\bm{x}_N)$ and $1:N$ are timesteps of the sequence for our cascade LSTM network. Suppose that the corresponding label of $\bm{X}$ is $\Delta{\bm{Y}}$. Note that $\Delta{\bm{Y}}$ denotes the total increment of $n$ timesteps. In a certain practical application, if the time step of the ground truth is $T$, the time step of fusion navigation could be $t=T/n(n>1)$. To achieve high frequency navigation, the prediction for each time step $t$ should be produced. We cascade $n$ LSTMs to simulate incremental changes of $n$ timesteps, and each LSTM is used to estimate the increment for one time step. In this study, his method is called $\theta $-increment learning which learns increments of variables using the constructed cascade LSTM network for obtaining high frequency estimation, and $\theta $ denotes the target variable to be estimated.

Given that $n$ LSTMs aim at learning the same relationship between their inputs and outputs, we let all these $n$ LSTMs share parameters to be learned. This shared mode facilitates the training of the entire cascade LSTM network, and each LSTM is called a shared LSTM (S\_LSTM) cell. We assume the input of the $i^{th}$ S\_LSTM is $\bm{X}_i=(\bm{x}_i,\bm{x}_{(i+1)},\cdots,\bm{x}_{(m+i-1)})$ and its output is ${\Delta{\bm{Y}_{i}}}$, where $1:m$ represent timesteps of each S\_LSTM and $N=m+n-1$. The final estimation of the cascade LSTM network is $\Delta{\widehat{\bm{Y}}}={\Delta{\bm{\widehat{Y}}_{1}}}+{\Delta{\bm{\widehat{Y}}_{2}}}+\cdots +{\Delta{\bm{\widehat{Y}}_{n}}}$, and $\oplus$ in Fig.~\ref{fig2} denotes the summation operation. The entire cascade LSTM network is trained using the Adam optimizer according to the $mse$ loss between the label $\Delta{\bm{Y}}$ and the predicted result $\Delta{\widehat{\bm{Y}}}$.

Thanks to the adaptive characteristics of LSTM, the cascade LSTM based $\theta $-increment learning method does not need end-to-end training data and the shared parameters are updated every $n$ timesteps in our study. Moreover, benefiting from these adaptive characteristics, the shared LSTM cell is capable to accurately predict the increment of one small time step. Therefore, the trained shared LSTM cell can obtain predictions with high frequency and high precision which could be higher than that of the training data.
Note that $\Gamma $ in Fig.~\ref{fig2} denotes the initialization operation. The usage of $\Gamma $ depends on the relationship between the target variable and time, which will be described in detail below.

\subsection{Cascade LSTM based Orientation and Position Estimation}
\label{PCasLSTM}

In the maglev haptic system, 6DOF navigation information (3DOF position and 3DOF orientation) of the magnetic stylus should be acquired for capturing users' interaction operation. In this work, visual-inertial navigation is implemented by using the presented cascade LSTM based $\theta $-increment learning method. Specifically, two cascade LSTM networks are separately trained for estimating the position and orientation of moving objects. Considering that IMUs are capable to obtain high-frequency sampling and the visual module can acquire high-precision positioning information \cite{tong2017magnetic}, the inertial data with high frequency is utilized as the input of our deep learning model and the ground truth of position is obtained by adopting the visual navigation method described in \cite{tong2017magnetic}. Note that the ground truth of orientation is calculated by using one inertial controller with a high-precision on board Digital Motion Processor (DMP).
%the on board Digital Motion Processor (DMP) with high-precision.
%one inertial controller which contains a built-in digital motion processor with high-precision.
%orientation calculation unit (embedded motion driver) with high-precision.

\subsubsection{Cascade LSTM based orientation estimation}

The cascade LSTM network used for orientation estimation is called OCasLSTM. For the orientation estimation, the acceleration and angular rate $\bm{I}$ acquired by IMUs are the input of OCasLSTM, and the total increment $\Delta{\bm{{Q}}}$ of $n$ shared LSTM cells for the orientation is the output of OCasLSTM, as shown in Fig.~\ref{fig2}(b). Because the input is the first-order derivative of orientation, the output of each shared LSTM cell in OCasLSTM exactly corresponds to the increment of orientation. Therefore, the orientation estimation does not need the initialization operation.

During the training, the $mse$ loss is utilized to update OCasLSTM, and $\Delta \bm{Q}$ is the summation of $n$ increments ($\Delta {\bm{q}_{1}},\Delta {\bm{q}_{2}},\cdots ,\Delta {\bm{q}_{n}}$) obtained from $n$ shared LSTM cells. In practical applications, only one shared LSTM cell is needed to predict the increment of orientation for one time step, and the current estimated orientation is the sum of the predicted increment and the orientation of the previous moment.

\subsubsection{Cascade LSTM based position estimation}

The cascade LSTM network used for position estimation is called PCasLSTM.
%Similarly, the acceleration and angular rate acquired by IMUs are regarded as the input $\bm{I}$ of PCasLSTM, and the total increment of $n$ components for the position is the output $\Delta{\bm{{P}}}$ of PCasLSTM.
Different from the orientation, the position is the double integration of the acceleration and angular rate. According to the kinematics theory, an initial velocity except for the acceleration and angular rate should be provided for calculating the increment of position. To tackle this issue, we introduce the initialization operation $\Gamma $ into PCasLSTM, as shown in Fig.~\ref{fig2}(c).

For the initialization operation $\Gamma $, the increment of position and its corresponding time are known. To obtain the initial velocity, the motion is assumed to be a certain state which can be uniform velocity, uniform acceleration, etc. In order to alleviate the impact of such an initialization operation, PCasLSTM is constructed by cascading $k$ CasLSTM units, and these units are trained simultaneously. For the first unit, the increment of each timestep obtained using the initialization operation $\Gamma $ and the inertial data sequence $\bm{I}_{1}, \bm{I}_{2}, ..., \bm{I}_{N}$ are contacted as its input. For each following unit, the estimated position from the previous CasLSTM unit and the corresponding inertial data sequence are contacted as its input. 
%Besides, these cascade LSTM units do not share parameters.

The output of the $i^{th}$ CasLSTM unit $\Delta \bm{P}_i$ is the summation of $n$ increments (i.e. $\Delta {\bm{p}_{1}},\Delta {\bm{p}_{2}},\cdots ,\Delta {\bm{p}_{n}}$) obtained from $n$ shared LSTM cells. Note that $n$ shared LSTM cells of each CasLSTM share parameters, while $k$ CasLSTM units of PCasLSTM have separate parameter configurations. During the training, multiple losses are used to update the entire PCasLSTM. In practical applications, we firstly perform the initialization operation using the trained PCasLSTM, and then the shared LSTM cell of the last cascade LSTM unit in PCasLSTM is used to predict the increment of position for one time step. The current estimated position is the sum of the predicted increment and the position of the previous moment.

\subsection{Cascade LSTM based Visual-Inertial Fusion Navigation}

After finishing the offline training of cascade LSTM based orientation and position estimation models, we use the trained models to provide accurate and real-time navigation for the maglev haptic interaction system. The concrete steps for accomplishing visual-inertial fusion navigation are as follows:

$\bullet$ Preprocessing: After receiving the visual and inertial data of the magnetic stylus, the AI service firstly preprocesses these data, such as formatting and normalization.

$\bullet$ Initialization: As described in Section \ref{PCasLSTM}, the shared LSTM cell of the last CasLSTM unit in PCasLSTM is used for position estimation. We firstly perform the entire PCasLSTM to obtain an accurate initial state for the shared LSTM cell.
%trained simultaneously. to Lay down a set of variables used to hold the current position and orientation, whose initial values are set by the beginning data of cameras and IMUs. The trained model should be initialized according to the actual initial motion.

$\bullet$ Estimation: This step can be divided into three cases:

a) If only raw acceleration and angular rate data $\bm{I}$ are received, predict increments of position and orientation.
b) If position data is received except from $\bm{I}$, predict the increment of orientation.
c) If orientation data is received except from $\bm{I}$, predict the increment of position.

$\bullet$ Update: If position and orientation data are received except from $\bm{I}$, update the location information (position and orientation) of the magnetic stylus. Otherwise, perform the ``Estimation" step and update the location information of the magnetic stylus by adding estimated increments and the location information of the previous moment together.

\section{Experimentation Results}

\subsection{System Tested}

\begin{figure}[htbp]
  \centering
  \includegraphics[width=8cm]{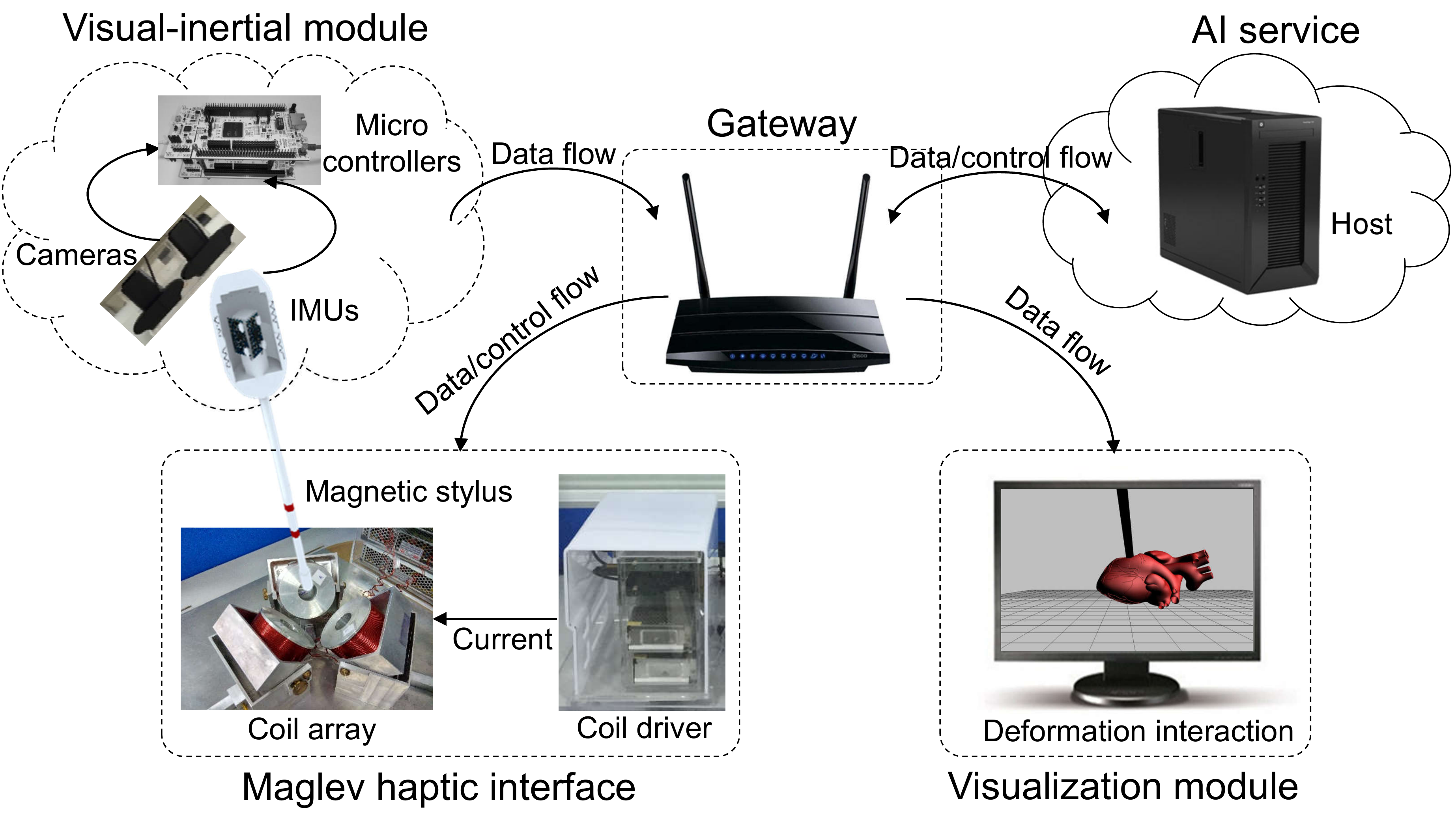}\\
  \caption{The system tested.}\label{fig3}
\end{figure}

In this section, to verify the cascade LSTM based visual-inertial navigation approach proposed by this article, we set up a system testbed in view of a maglev haptic interaction application. The maglev haptic interaction system is composed of a visual-inertial navigation module, a maglev haptic interface \cite{tong2016novel,tong2017magnetic}, an AI service and a visualization module. The visual-inertial navigation module is used to obtain visual and inertial data. The maglev haptic interface \cite{tong2016novel,tong2017magnetic} includes a magnetic stylus, a coil array and a coil driver module, and it provides haptic feedback in the haptic interaction application. The AI service performs visual-inertial navigation using deep learning models. Besides, the visualization module is utilized to show virtual scenes. %We use the maglev haptic interface to perform the interactive heart deformation simulation. 
In summary, the specific haptic interaction process is as follows.
When users move the magnetic stylus in the operation workspace of the maglev haptic interface, visual and inertial data collected or calculated by the visual-inertial module are sent to the AI service which performs cascade LSTM based visual-inertial navigation for haptic interaction. The visualization module displays the virtual heart deformation model in real time according to the navigation information, and meanwhile, the maglev haptic interface provides the corresponding haptic feedback to users.
%Red markers embedded in the a magnetic stylus is used for visual navigation and IMUs is used for inertial navigation, as shown in Fig.~\ref{fig3}.

As shown in Fig.~\ref{fig3}, the visual-inertial module, AI service, maglev haptic interface and visualization module communicate under the same LAN in real time. Two IMUs (MPU6050) are fixed on the back end of the magnetic stylus. One is used to gather the raw acceleration and angular rate data, and the other one is used for providing orientation angles. Cameras capture the position information of the magnetic stylus by tracking red markers. Our cascade LSTM based visual-inertial navigation approach can export the position and orientation of the magnetic stylus at a frequency of 200 Hz in the heart deformation haptic interaction application.
%All the data is sent to the host computer of the virtual scene in sync time at different frequencies.

The specific experimental process is as follows: First, we performed interactive heart deformation simulation on the system tested and collected 30,000 data. The collected data included acceleration and angular rate, and the corresponding position and orientation data, and their sampling frequency were 200Hz, 20Hz and 100Hz, respectively. Then, we divided the collected data into training set, validation set and testing set with the ratio of 8:1:1. Cascade LSTM based position and orientation estimation models were trained on the training set. After that, trained models were evaluated on the testing set.  %The frequency and accuracy of fusion navigation can be told by both vision and touch.

\subsection{Experimental Results and Analysis}

\begin{table}[ht]
\centering
\caption{The MAE between the predicted and actual orientation data.}\label{table:1}
\begin{center}
\renewcommand{\multirowsetup}{\centering}
\begin{tabular}{p{1.2cm}<{\centering}|p{1cm}<{\centering}|p{1cm}<{\centering}|p{1cm}<{\centering}|p{1cm}<{\centering}|p{1cm}<{\centering}}
\hline
Orientation & ratio=2 & ratio=4 & ratio=6 & ratio=8 & ratio=10    \\
\hline
Pitch ($^{\circ }$)  & 0.0086  & 0.0115  & 0.0136 & 0.0166 & 0.0198 \\
Roll ($^{\circ }$)   & 0.0092  & 0.0113  & 0.0143 & 0.0173 & 0.0200  \\
Yaw ($^{\circ }$)    & 0.0050  & 0.0060  & 0.0072 & 0.0080 & 0.0105  \\
\hline
\end{tabular}
\end{center}
\end{table}

In order to verify the performance of our cascade LSTM based $\theta $-increment learning method, we trained five OCasLSTM models, and the increase ratio of the frequency were 2, 4, 6, 8, and 10, respectively. To make sure fairness, the testing data were not used to train these models. Table~\ref{table:1} demonstrates the mean absolute error (MAE) between the predicted and actual orientation angles (pitch, roll and yaw).

\begin{figure}[htbp]
  \centering
  \includegraphics[width=9.3cm]{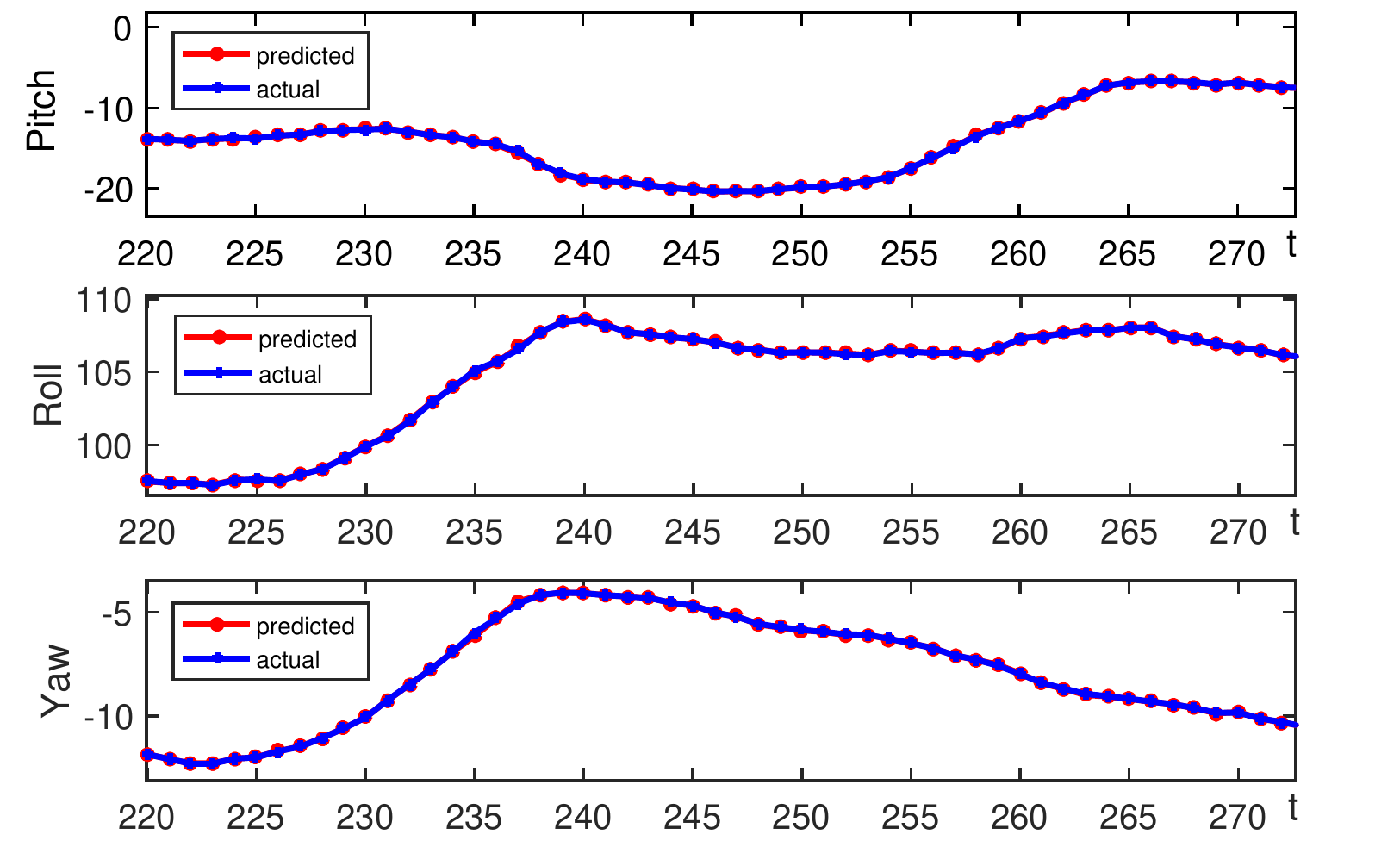}\\
  \caption{Comparison between the predicted and actual orientation angular.}\label{fig4}
\end{figure}

From Table~\ref{table:1}, we can see although the MAE value between the predicted and actual orientation data becomes higher as the ratio increases, the max MAE value is less than 0.02$^{\circ }$. Besides, Fig.~\ref{fig4} shows the comparison between the predicted orientation data of the OCasLSTM with the increase ratio of 10 and the actual orientation data, and the predicted results are very close to the actual ones. From Table~\ref{table:1} and Fig.~\ref{fig4}, it can be seen that the presented cascade LSTM based $\theta $-increment learning method is promising.

\begin{figure}[htbp]
  \centering
  \includegraphics[width=9.2cm]{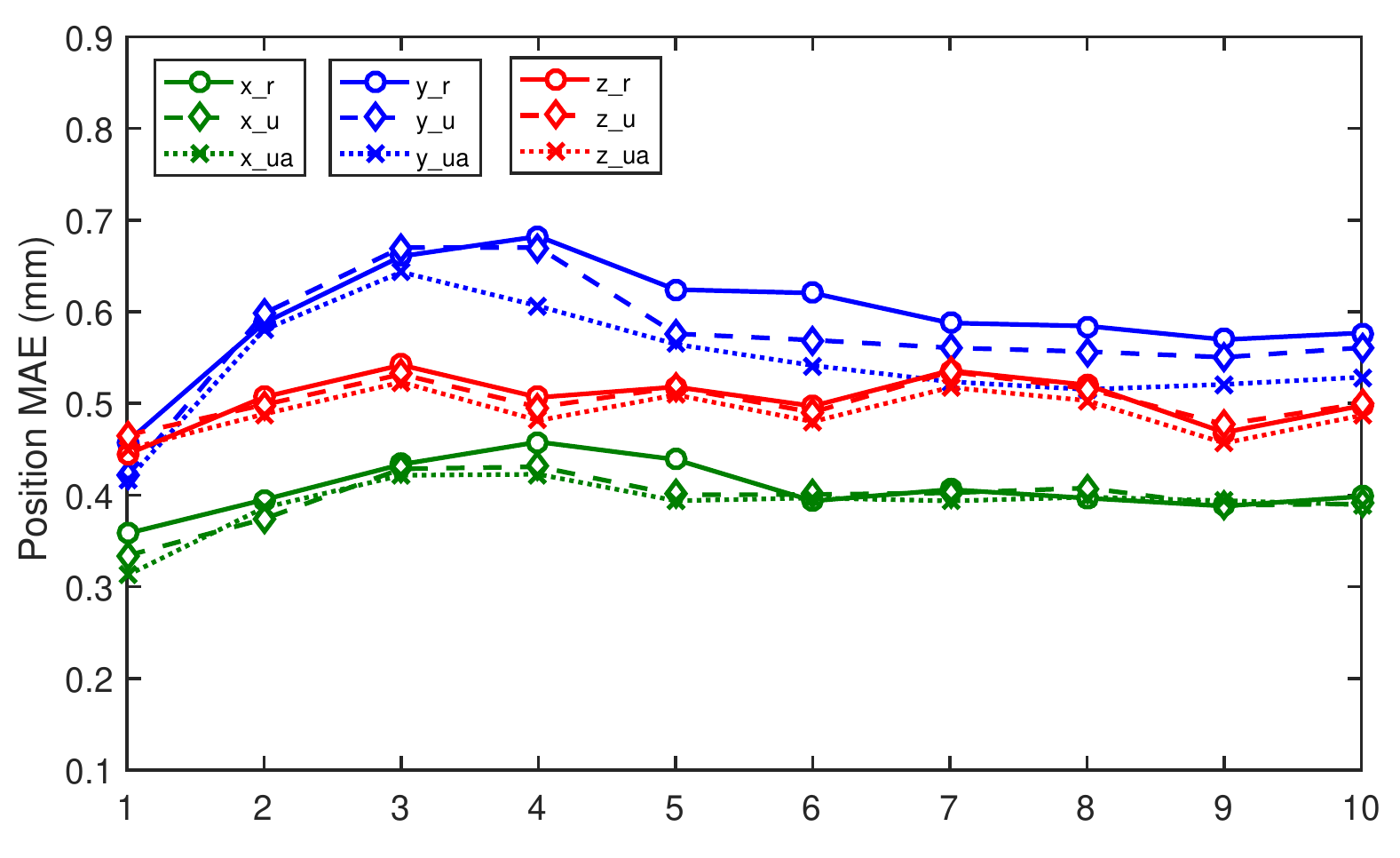}\\
  \caption{The MAE between the predicted and actual data of ten CasLSTM units for three initialization methods.}\label{fig5}
\end{figure}

Then, we trained three PCasLSTM models with ten CasLSTM units by using three different initialization methods (uniform velocity--u, uniform acceleration--ua and random--r). Fig.~\ref{fig5} shows the mean absolute error between the predicted and the actual position data of these ten CasLSTM units for three initialization methods. The mean absolute error of three initialization methods is very close, demonstrating the robust of PCasLSTM. Furthermore, the mean absolute error of ten cascade LSTM units for three initialization methods is less than 1mm, showing high accuracy of our approach.

\section{Conclusion}

In this work, we focus on the problem of how to improve the navigation frequency and stability while maintaining high-precision and the lightweight design for maglev haptic interaction. To achieve this goal, we present a cascade LSTM based $\theta $-increment learning method which is utilized to construct two separate cascade LSTM networks for accomplishing position and orientation estimation. This proposed cascade LSTM based visual-inertial navigation approach can yield position and orientation estimation of moving objects for a small time step, thus it can achieve high frequency navigation. Furthermore, the accuracy of our approach was verified by building a testbed. In future studies, we will research high-precision haptic rendering methods based on the proposed visual-inertial navigation approach, and extend the navigation approach to other applications.

%\begin{figure*}[htbp]
%  \centering
%  \includegraphics[width=15cm]{Figure5.png}\\
%  \caption{The left part is our magnetic levitation haptic device and the right part depicts its overall design.}\label{fig5}
%\end{figure*}

%% use section* for acknowledgment
%\section*{Acknowledgment}
%%
%This work is supported by the Shenzhen Science and Technology Program (Project No.JCYJ20160429190300857) and the National Natural Science Foundation of China (Grant No. 61802385).
%The work was supported by the National Natural Science Foundation of China under Grant No. 61373107, Wuhan Science and Technology Program under Grant No. 2016010101010022.

\ifCLASSOPTIONcaptionsoff
  \newpage
\fi



%\bibliographystyle{IEEEtran}

\bibliographystyle{IEEEtran}
%% argument is your BibTeX string definitions and bibliography database(s)
\bibliography{references}

\begin{thebibliography}{10}
\providecommand{\url}[1]{#1}
\csname url@samestyle\endcsname
\providecommand{\newblock}{\relax}
\providecommand{\bibinfo}[2]{#2}
\providecommand{\BIBentrySTDinterwordspacing}{\spaceskip=0pt\relax}
\providecommand{\BIBentryALTinterwordstretchfactor}{4}
\providecommand{\BIBentryALTinterwordspacing}{\spaceskip=\fontdimen2\font plus
\BIBentryALTinterwordstretchfactor\fontdimen3\font minus
  \fontdimen4\font\relax}
\providecommand{\BIBforeignlanguage}[2]{{%
\expandafter\ifx\csname l@#1\endcsname\relax
\typeout{** WARNING: IEEEtran.bst: No hyphenation pattern has been}%
\typeout{** loaded for the language `#1'. Using the pattern for}%
\typeout{** the default language instead.}%
\else
\language=\csname l@#1\endcsname
\fi
#2}}
\providecommand{\BIBdecl}{\relax}
\BIBdecl

\bibitem{chen2018wearable}
M.~Chen, J.~Zhou, G.~Tao, J.~Yang, and L.~Hu, ``Wearable affective robot,''
  \emph{IEEE Access}, vol.~6, pp. 64\,766--64\,776, 2018.

\bibitem{elbamby2018toward}
M.~S. Elbamby, C.~Perfecto, M.~Bennis, and K.~Doppler, ``Toward low-latency and
  ultra-reliable virtual reality,'' \emph{IEEE Network}, vol.~32, no.~2, pp.
  78--84, 2018.

\bibitem{berkelman2012co}
P.~Berkelman, M.~Miyasaka, and J.~Anderson, ``Co-located 3d graphic and haptic
  display using electromagnetic levitation,'' in \emph{Haptics Symposium
  (HAPTICS), 2012 IEEE}.\hskip 1em plus 0.5em minus 0.4em\relax IEEE, 2012, pp.
  77--81.

\bibitem{pedram2017torque}
S.~A. Pedram, R.~L. Klatzky, and P.~Berkelman, ``Torque contribution to haptic
  rendering of virtual textures,'' \emph{IEEE transactions on haptics},
  vol.~10, no.~4, pp. 567--579, 2017.

\bibitem{tong2016novel}
Q.~Tong, Z.~Yuan, M.~Zheng, W.~Zhu, G.~Zhang, and X.~Liao, ``A novel magnetic
  levitation haptic device for augmentation of tissue stiffness perception,''
  in \emph{Proceedings of the 22nd ACM Conference on Virtual Reality Software
  and Technology}.\hskip 1em plus 0.5em minus 0.4em\relax ACM, 2016, pp.
  143--152.

\bibitem{tong2017magnetic}
Q.~Tong, Z.~Yuan, X.~Liao, M.~Zheng, T.~Yuan, and J.~Zhao, ``Magnetic
  levitation haptic augmentation for virtual tissue stiffness perception,''
  \emph{IEEE Transactions on Visualization and Computer Graphics}, 2017.

\bibitem{weiss2011real}
S.~Weiss and R.~Y. Siegwart, ``Real-time metric state estimation for modular
  vision-inertial systems,'' in \emph{Robotics and automation, 2011 IEEE
  International Conference on}.\hskip 1em plus 0.5em minus 0.4em\relax IEEE,
  2011, pp. 4531--4537.

\bibitem{mourikis2007multi}
A.~I. Mourikis and S.~I. Roumeliotis, ``A multi-state constraint kalman filter
  for vision-aided inertial navigation,'' in \emph{Robotics and automation,
  2007 IEEE international conference on}.\hskip 1em plus 0.5em minus
  0.4em\relax IEEE, 2007, pp. 3565--3572.

\bibitem{leutenegger2015keyframe}
S.~Leutenegger, S.~Lynen, M.~Bosse, R.~Siegwart, and P.~Furgale,
  ``Keyframe-based visual--inertial odometry using nonlinear optimization,''
  \emph{The International Journal of Robotics Research}, vol.~34, no.~3, pp.
  314--334, 2015.

\bibitem{qin2018vins}
T.~Qin, P.~Li, and S.~Shen, ``Vins-mono: A robust and versatile monocular
  visual-inertial state estimator,'' \emph{IEEE Transactions on Robotics},
  vol.~34, no.~4, pp. 1004--1020, 2018.

\bibitem{chen2018label}
M.~Chen, Y.~Hao, K.~Lin, Z.~Yuan, and L.~Hu, ``Label-less learning for traffic
  control in an edge network,'' \emph{IEEE Network}, vol.~32, no.~6, pp. 8--14,
  2018.

\bibitem{chen2017deep}
M.~Chen, X.~Shi, Y.~Zhang, D.~Wu, and M.~Guizani, ``Deep features learning for
  medical image analysis with convolutional autoencoder neural network,''
  \emph{IEEE Transactions on Big Data}, DOI: 10.1109/TBDATA.2017.2717439, 2017.

\bibitem{greff2017lstm}
K.~Greff, R.~K. Srivastava, J.~Koutn{\'\i}k, B.~R. Steunebrink, and
  J.~Schmidhuber, ``Lstm: A search space odyssey,'' \emph{IEEE transactions on
  neural networks and learning systems}, vol.~28, no.~10, pp. 2222--2232, 2017.

\bibitem{Delmerico2018A}
J.~Delmerico and D.~Scaramuzza, ``A benchmark comparison of monocular
  visual-inertial odometry algorithms for flying robots,'' in \emph{IEEE
  International Conference on Robotics and Automation}, 2018.

\bibitem{clark2017vinet}
R.~Clark, S.~Wang, H.~Wen, A.~Markham, and N.~Trigoni, ``Vinet: Visual-inertial
  odometry as a sequence-to-sequence learning problem.'' in \emph{AAAI}, 2017,
  pp. 3995--4001.

\end{thebibliography}

\end{document}